\documentclass[letterpaper]{JHEP3}

\usepackage{epsfig}
\usepackage{slashed}

\title{Missing energy in black hole production and decay at the Large
       Hadron Collider}

\author{
Douglas M. Gingrich\\ 
Centre for Particle Physics, Department of Physics, University of
Alberta, Edmonton, AB T6G 2G7 Canada\\
TRIUMF, Vancouver, BC V6T 2A3 Canada\\ 
E-mail: \email{gingrich@ualberta.ca}
}

\abstract{
Black holes could be produced at the Large Hadron Collider in TeV-scale
gravity scenarios.
We discuss missing energy mechanisms in black hole production and decay  
in large extra-dimensional models.
In particular, we examine how graviton emission into the bulk could
give the black hole enough recoil to leave the brane.
Such a perturbation would cause an abrupt termination in Hawking
emission and result in large missing-energy signatures. 
}


\preprint{}

\received{December 9, 2006}
\accepted{December 9, 2006}

\begin{document}

\section{Introduction\label{sec1}}

Models of large extra dimensions allow the fundamental scale of gravity
to be as low as the electroweak
scale~\cite{Arkani98,Antoniadis,Arkani99}.   
If the gravity scale is as low as a TeV, black holes could be produced
in collisions at the Large Hadron Collider (LHC).
Once formed, the black hole will decay by emitting Hawking
radiation~\cite{Hawking75}. 
In these models, our universe has a domain wall structure, the brane,
that is embedded in a higher-dimensional bulk spacetime. 
In the class of models we will consider, the Standard Model particles
are confined to the brane while the graviton is allowed to exist in all
the dimensions.
Graviton emission would be observed as missing energy and, if the recoil
given to the black hole is significant, may result in the black hole
leaving the brane during the decay process. 

Numerous authors have estimated an enormous black hole production rate
of about $10^7$ per year at the LHC. 
This large rate is due to an anticipated parton-parton cross section
that rises geometrically with increasing parton-parton centre of mass
energy, and the assumption that all partons partake equally in black
hole formation.    
Only the steeply falling parton's momentum distribution in the proton
keep the cross section reasonably finite.

There are several ways to reduce the black hole cross section yet still
allow black holes to be produced at the LHC.
The classical parton cross section will probably not hold at
parton-parton collision energies near the fundamental Planck scale
$M_\mathrm{P}$.  
It is possible that the LHC will operate in the regime in which the
effects of quantum gravity can not be ignored.
In the quantum regime, the black hole may become stringy (string ball) and
have a different cross section energy
dependence~\cite{Dimopoulos02,Cheung02p}. 
Even if we are well above the Planck scale and clear of the effects of
quantum gravity, it has recently been pointed out that the contribution
to the stress energy tensor in Einstein's equations due to charged, and
perhaps coloured, partons will prevent all partons in the proton from 
contributing equally to black hole
production~\cite{Yoshino06a,Gingrich06c}.  
Some charged partons many not contribute at all under certain kinematic 
conditions and regions of higher-dimensional parameter space. 

Perhaps the largest uncertainty in the classical black hole cross
section picture is due to gravitational radiation during black hole
formation. 
Apparent-horizon studies give lower bounds on the amount of energy
that could be trapped behind the horizon during back hole
formation~\cite{Yoshino05a,Gingrich06a}. 
Although it is not known how much of the lower bound is due to the
apparent-horizon technique, results in four dimensions indicate that
significant radiation could be emitted~\cite{Eath93}.  
Assuming the black hole cross section is a function of the black hole
mass, initial radiation could significantly lower the production cross 
section. 
Even if the black hole is considered to have formed before the radiation
is emitted, the gravitational radiation will result in missing energy,
and the back hole will effectively have a lower mass before it begins to
be detectable by its Hawking radiation on the brane.  

It has long been argued that if black holes are produced at the LHC,
they will give rise to spectacular decay
signatures~\cite{Banks,Dimopoulos01,Giddings01a}.
The Hawking evaporation of these very hot black holes is expected to
generate high-multiplicity, almost spherical events with several very
high-energy jets, high-$p_\mathrm{T}$ leptons, and possibly even exotic 
particles~\cite{Landsberg02}. 
Black holes could also be identified by their large missing-energy
signatures from high-energy neutrinos emitted in the evaporation
process.  
However, large missing-energy events need to be removed from the data
sample, or well understood, to enable an accurate reconstruction of the
black hole mass~\cite{Harris05a,Tanaka}.  
It is the measurement of the black hole mass and event rate that will
allow us to infer the Planck scale, and possibly, the number of extra
dimensions. 

There are additional missing-energy signatures that could make the
reconstruction of black holes extremely difficult.
It is not know how the evaporation process will end.
The possibility of a final black hole remnant with mass of the order of
the Planck scale has been studied~\cite{Koch05}.
Either this remnant is charged and ionizing like a particle, in
which case it will need to be detected, or more likely~\cite{Casher}, it
will be neutral and possibly not detectable.

A final possibility giving missing energy, that we will examine, is that
the black hole will be perturbed during the decay process and could
leave the brane. 
If the black hole escapes, it will result in large missing energy.
It is unlikely that the black hole would leave the brane at a
particular mass value, but it will probably be a stochastic process
depending on the initial black hole mass and the history of the
evaporation process. 
Such a phenomena may assist in black hole identification, but it might
make accurate kinematic reconstruction of the black hole mass difficult.  
Hawking radiation is thermal and leads to unique democratic signatures,
but it is not clear if these signatures could be identified for those
events in which the black hole leaves the brane.

With all these missing-energy signatures, we argue that it may be
difficult to conclusively detect black holes if they are produced at the
LHC.   
Since a black hole is not a mass resonance, we will not know \textit{a
priori} what initial energy went into the black hole formation; energy
will be lost during the production process.  
Once formed, black hole states may be difficult to accurately
reconstruct event by event from the final-state particles because of
widely varying missing energy in the decay process.     

In this paper, we discuss missing energy due to mechanisms that are not
usually mentioned in the literature on TeV-scale black hole production
and decay. 
We concentrate our study on missing energy from Schwarzschild black
holes that leave the Standard Model brane during the Hawking evaporation 
process. 
An outline of this paper is as follows.
In Sec.~\ref{sec2} we discuss qualitatively black hole production and
decay with particular emphasis on the graviton radiation processes.
In Sec.~\ref{sec3} we review the Hawking evaporation process in higher
dimensions with particular attention to the graviton mode.
We briefly review a mechanism for black hole escape from the brane and a 
model of the binding potential in Sec.~\ref{sec4}.
In Sec.~\ref{sec5} we describe our model and simulation.   
Graviton emission probabilities, particle multiplicities, and the
probability for the black hole to leave the brane, along with missing
energy distributions and their effect on the black hole mass and cross
section are examined in Sec~\ref{sec6}. 
In Sec.~\ref{sec7} we conclude with a discussion.

\section{Black hole production and decay\label{sec2}}

Black hole production and decay can be thought of as evolving according
to a series of distinct phases.   
In the production phase, the gravitational fields of the relativistic 
particles producing the black hole are approximately localized to narrow
longitudinal shock waves and spacetime is flat before the collision.
At the instance of collision, the two shock waves pass through one
another, and interact nonlinearly by shearing and focusing.
After the collision, the two shocks continue to interact nonlinearly
with each other and spacetime within the future lightcone of the
collision becomes highly curved.
A complex-shaped event horizon forms which quickly collapses down to
a more regular-shaped apparent horizon by the emission of gravitational 
waves into the bulk space. 
Not all the energy in the two-particle collision is trapped behind the
horizon and the collision process can be considered to be inelastic.
The effect of inelasticity is to reduce the black hole mass and thus
cross section, but otherwise does not give observable signatures.
The black hole produced may have any gauge or angular momentum quantum
numbers arising from the two initial partons.

According to the no-hair theorem~\cite{Hawking05}, the resulting
asymmetry and moments due to the violent production process are radiated
away by gravitons into the bulk until a Kerr-Newman stationary solution
is formed, which is characterized by only its mass, angular momentum,
and local charges (electric charge and probably colour charge).   
For excited black holes produced in four dimensions by neutral
relativistic particles, 16\% of the total energy is lost in this balding
phase~\cite{Eath93}.

Due to conservation of angular momentum, the angular momentum of the
formed black hole can only vanish completely for central collisions with
zero impact parameter.  
In the general case of an impact parameter, there will be an angular
momentum.
Black holes are expected to be produced in high angular-momentum states
from particle collisions above the Planck scale.
It is anticipated that they will spin down by Hawking evaporation very
rapidly to Reissner-Nordstr\"{o}m static solutions by the emission of
high-spin state particles.   
In four dimensions, the half-life of the spin down phase is 7\% of the
black hole lifetime, and about 25\% of the mass is lost during this
spin-down phase~\cite{Page76b}.
The various emissivities are enhanced by a factor of about 35 to almost
100 as the number of dimensions increase; this factor increases by 3 to
6 as the angular momentum of the black hole increases~\cite{Casals06}. 
In higher dimensions, black holes also tend to lose their angular
momentum at the early stage of evolution.
However, black holes can still have a sizable rotation parameter after
radiating half their mass.
Typically more than 70\% to 80\% of the black hole's mass is lost during
the spin-down phase~\cite{Ida06}.

The Schwarzschild evaporation phase (Hawking evaporation of a
non-rotating black hole) is the most well studied.
A black hole of a particular mass is characterized by a Hawking
temperature and, as the decay progresses, the black hole mass falls and
the temperature rises.
Thermal radiation is thought to be emitted by black holes due to quantum 
effects. 
Grey-body factors modify the spectrum of emitted particles from that
of a perfect thermal black body~\cite{Page76a} and quantify the
probability of transmission of the particles through curved spacetime
outside the horizon.  
At high energies, the shape of the spectrum is like that of a black body
while at low energies the behaviour of the grey-body factors is
spin-dependent and also depends on the number of dimensions.

Baryon number ($B$) and lepton number ($L$) do no have to be conserved in
black hole decay. 
However, it is widely believed that $B-L$ is conserved, which would help
bind the black hole to the brane.

Since Hawking radiation allows black holes to lose mass, they could
evaporate, shrink, and ultimately vanish.   
A black hole can not decay down to nothing without the loss of
information. 
Another possibility is that a black hole could leave a sub-Planckian
remnant. 
The final fate of a black hole is unknown since quantum gravity will
become important as the black hole mass approaches the Planck scale.

\subsection{Gravitational radiation during black hole formation\label{sec2.1}}

It is not known at what instance a black hole would form in particle
collisions in low-scale gravity scenarios. 
During the formation process, significant amounts of gravitational
radiation would probably be emitted.
Likewise, significant gravitational radiation could be emitted near the 
threshold for black hole formation even if the black hole does not form.

Pretorius and Khurana~\cite{Pretorius07} have performed numerical
studies in four dimensions of black hole mergers and unstable circular
orbits for a class of equal-mass, non-rotating, non-circular binary
black hole systems in general relativity.
They find evidence of an approximate correspondence between
near-threshold evolution of geodesics and generic binary merger. 
They applied this correspondence to the threshold for black hole
production in particle collisions of high energy.
The merger of two black holes is thought to be equivalent to black hole
formation in particle collisions of sufficiently high energy where the
classical general relativistic description holds. 
Ideally we would like to know the threshold impact parameter below which
a black hole forms and the energy radiated as a function of impact
parameter. 
Pretorius and Khurana find that at threshold it is possible that
essentially all the kinetic energy is radiated as gravitational waves
and that there is still significant energy loss to gravitational waves
for impact parameters up to almost twice the critical value for black
hole formation. 
Although these studies are in four dimensions, no counter arguments have
been made to indicate that they are not applicable in higher dimensions. 

Since black hole production at the LHC mostly occurs from quark-quark
interactions, most of the produced black holes will have fractional
electric change and colour.
These quantum numbers will make it difficult for the black hole to leave
the brane before this hair can be shed during the balding phase.

\subsection{Gravitational radiation after black hole formation\label{sec2.2}}

During the balding phase, the black hole is considered to exist.
The black hole settles down into a Kerr-Newman solution by eliminating
its moments in gravitational radiation.
Although Kerr-Newman solutions are unique in four dimensions this is not
the case in higher dimensions~\cite{Elvang,Ida03d}.
Indeed, black hole Saturn solutions have been found in five dimensions
and are anticipated to exist in other higher dimensions.

It will be difficult to obtain experimental information about black hole
formation and the balding phase; all emitted radiation is undetectable 
gravitational waves formed from partons of unknown initial energies.
Gravitational radiation will result in lowering the black hole mass
before the spin-down phase begins.
Thus measurement of the cross section may have to be corrected for
gravitational radiation by theory and modeling in order to obtain the
true mass dependence.  
This situation is not dissimilar to electromagnetic or QCD initial-state 
radiation in which the radiation can not be detected.
To make the correction more difficult, the black hole is not a particle
with a definite mass and the amount of radiation has not yet been
formulated. 

Black holes are expected to be highly rotating when produced in particle
collisions.
A black hole can thus exhibit superradiance in its decay.
This enhances the emission of higher spin-state particles possibly
making the emission of gravitons a dominant effect.
In four dimensions, Page~\cite{Page76b} showed that the probability of 
emission of a graviton by an extremely rotating black hole is 100 times
higher than the probability of emission of a photon or a neutrino. 
In four dimensions, graviton emission, which is suppressed for small
rotations, rapidly increases with angular momentum, but the angular
momentum is restricted to $J < M^2$.
In higher dimensions, there is no upper bound on $J$ and so graviton
emission could dominate the evaporation process for rotating black
holes. 

Since gravitons are not bound to the brane, most would radiate into the
bulk. 
Although black holes produced at the LHC would initially have no
components of angular momentum in the higher dimensions, the bulk 
components would soon become nonzero due to graviton emission.
A rotating black hole could lose its bulk components of rotation by
interacting with the brane or emitting further Hawking quanta into the
bulk. 
Emission of gravitons into the bulk during the spin-down phase could
thus strongly perturb the system, possibly causing the black hole to
leave the brane.  

\section{Hawking evaporation\label{sec3}}

Hawking radiation provides distinct experimental signatures that may
allow discrimination between gravitational events and other perturbative
non-gravitational physics. 
For an uncharged, non-rotating black hole, the decay spectrum per degree
of freedom $s$ is described by  

\begin{equation} \label{eq1}
\frac{dN^{(s)}(\omega)}{dt d\omega} = \frac{1}{2\pi}
\frac{\Gamma^{(s)}(\omega)}{\exp(\omega/T_\mathrm{H}) \mp 1}\, , 
\end{equation}

\noindent
where $\omega$ is the energy of the emitted quanta, $T_\mathrm{H}$ is
the temperature of the black hole, and $\Gamma^{(s)}(\omega)$ is the
grey-body factor for mode $s$.
The last term in the denominator is a spin-statistics factor which is
$-1$ for bosons or $+1$ for fermions.
Equation~(\ref{eq1}) refers to individual degrees of freedom not
elementary particles. 
However, it can be used to determine the decay spectrum for a particular
particle by summing over the number of degrees of freedom. 

It has been thought that the majority of the energy in Hawking radiation
is emitted into Standard Model particles, but a small amount is also
emitted into gravitons~\cite{Emparan00}.  
A common argument in support of this claim is that fewer particles are
emitted in the bulk than on the brane; only the graviton is emitted in
the bulk, whereas all the Standard Model fields are emitted on the
brane.   
However, the emission rate per degree of freedom of the graviton in the
$D$-dimensional bulk could be higher than that of the four-dimensional
brane modes. 
It is now thought that the probability of emitting spin-two quanta in
high dimensions is substantial~\cite{Cavaglia03c}. 

In four dimensions, the graviton power loss is negligible compared to
the loss in Standard Model channels.
The Standard Model emissivities should not change much in higher
dimensions while the graviton emissivity is expected to be higher in
higher dimensions due to the increase in number of helicity states.   
In four dimensions, gravitational waves have two possible helicities.
In $D$ dimensions, the number of helicities is 

\begin{equation} \label{eq2}
\mathcal{N} = \frac{D(D-3)}{2}\, .
\end{equation}

\noindent
Thus in 11 dimensions the number of graviton helicity states reaches
44. 
In addition, the total power radiated in gravitons increases more
rapidly than the power radiated in lower-spin fields as the number of
dimensions increases. 
This is due to the increase in the multiplicity of the tensor
perturbations~\cite{Cavaglia03c,Park06,Creek06}. 

\subsection{Degrees of freedom\label{sec3.1}}

We assume the particle content at trans-Planckian energies will be the
minimal $U(1)\times SU(2)\times SU(3)$ Standard Model with three
families and one Higgs field.  
The number of degrees of freedom (dof) for each particle is given by 

\begin{equation} \label{eq3}
\mathrm{dof} = n_Q \times n_S \times n_\mathrm{F} \times n_\mathrm{C}\,
, 
\end{equation}

\noindent
where $n_Q$ is the number of charge states, $n_S$ the number of spin
polarizations, $n_\mathrm{F}$ the number of flavours, and $n_\mathrm{C}$
the number of colours.
Not all these degrees of freedom apply to each type of particle.
For massive gauge bosons one of their degrees of freedom comes from the
Higgs mechanism.
This means for each massive gauge boson there is one spin-0 degree of
freedom and two spin-1 degrees of freedom.
The number of degrees of freedom for each Standard Model particle is
shown in Table~\ref{dof}. 
The number of degrees of freedom (helicities) for the graviton will be
accounted for in the emissivity since it depend on the number of
dimensions.  

\TABULAR{|l|cccc|c|}
{\hline
Particle Type     & Charge & Spin   & Flavour & Colour & dof \\
                  & States & States & States  & States &     \\ \hline
quarks            & 2      & 2      & 6       & 3      & 72  \\
charged leptons   & 2      & 2      & 3       &        & 12  \\
neutrinos         & 2      & 1      & 3       &        & 6   \\
gluons            & 1      & 2      &         & 8      & 16  \\
photon            & 1      & 2      &         &        & 2   \\
Z boson           & 1      & 3      &         &        & 3   \\
W bosons          & 2      & 3      &         &        & 6   \\
Higgs             & 1      &        &         &        & 1   \\
graviton          & 1      &        &         &        & 1   \\
\hline}
{Number of degrees of freedom (dof) of the Standard Model
particles.
\label{dof}}

The picture of a massless graviton propagating in $D$ dimensions and the
picture of massive Kaluza-Klein (KK) gravitons propagating in four
dimensions are equivalent.
The $D(D-3)/2$ helicity states of the massless graviton in $D$
dimensions can be decomposed into KK helicity states: 1 scalar state,
$(D-3)$ vector states, and $(D-4)(D-1)/2$ tensor states. 

The KK picture allows one to write down an effective theory of
interactions of KK gravitons with Standard Model particles. 
This effective theory will breakdown above the Planck scale where black 
holes are active. 
The gravitons propagate in the extra dimensions and can decay into
ordinary particles only by interacting with the brane, and therefore
with a rate suppressed by $1/M_\mathrm{P}^2$.
The KK excitations of the graviton have the same very weak coupling to
the Standard Model fields as their massless zero mode.
This is because the graviton decaying weakly to ordinary matter is not
compensated by the large phase space of KK states.
We will thus assume the KK gravitons behave like massive,
non-interacting, stable particles, and that this assumption also holds
in the trans-Planckian region.
KK states can be produced in Standard Model particle collisions with a
reasonable strength.
Like all previous work on Hawking evaporation, we ignore the
interactions of all particles, including the KK gravitons.
We will also ignore possible light Nambu-Goldstone fields related to the
brane dynamics.   

\subsection{Emission spectra and probabilities\label{sec3.2}}

In the following, we will need the relative emission rates and the
shapes of the emission spectra.
The flux spectrum (number of particles emitted per unit time) is given
by\footnote{Throughout this paper we use $D$ to represent the total
number of spacetime dimensions, but in this section we use the common
convention of $n$ to represent the number of extra space dimensions:
$D = (n+4)$.} 

\begin{equation} \label{eq4}
\frac{dN^{(s)}(\omega)}{dt} = \sum_l \sigma_{n,l}^{(s)}(\omega)
\frac{1}{\exp(\omega/T_\mathrm{H})\mp 1} \frac{d^{n+3}k}{(2\pi)^{n+3}}\, ,
\end{equation}

\noindent
where

\begin{equation} \label{eq5}
\sigma_{n,l}^{(s)}(\omega) = \frac{2^n\pi^{(n+1)/2}\Gamma[(n+1)/2]}{n!\;
\omega^{n+2}} \frac{(2l+n+1)(l+n)!}{l!} \left|
\mathcal{A}_l^{(s)}(\omega)\right|^2 
\end{equation}

\noindent
is the grey-body factor for an $(n+4)$-dimensional Schwarzschild black
hole. 
The quantity $\sigma_{n,l}^{(s)}(\omega)$ is alternatively called the
partial absorption cross section.
It is the absorption (or transmission) probability for a scalar particle 
propagating in the brane background.
For a black body, $\sigma_{n,l}^{(s)}(\omega)$ is just a constant
representing the area of the emitting body. 
The absorption coefficient $\mathcal{A}_l^{(s)}(\omega)$ is not the
grey-body factor.    
Equation~(\ref{eq4}) is for a non-rotating, non-charged black hole.
For a rotating or charged black hole, the argument of the exponential is 
replaced by a more general expression.

For massless particles, we can integrate over the solid angle to obtain
the flux spectrum 

\begin{equation} \label{eq6}
\frac{dN^{(s)}(\omega)}{dt} = \sum_l \mathcal{N}_l
\left|\mathcal{A}_l^{(s)}(\omega)\right|^2
\frac{1}{\exp(\omega/T_\mathrm{H})\mp 1} \frac{d\omega}{2\pi}\, , 
\end{equation}

\noindent
where 

\begin{equation} \label{eq7}
\mathcal{N}_l = \frac{(2l+n+1)(l+n)!}{l!(n+1)!}
\end{equation}

\noindent
is the multiplicity of scalar modes for partial wave $l$.
The sum in eq.~(\ref{eq6}) can be removed by writing

\begin{equation} \label{eq8}
\frac{dN^{(s)}(\omega)}{dt} =
\frac{\Gamma^{(s)}(\omega)}{\exp(\omega/T_\mathrm{H})\mp 1}
\frac{d\omega}{2\pi}\, ,  
\end{equation}

\noindent
where

\begin{equation} \label{eq9}
\Gamma^{(s)}(\omega) = \sum_l \mathcal{N}_l \left|
\mathcal{A}_l^{(s)}(\omega) \right|^2\, . 
\end{equation}

\noindent
This result is identical to the previous expression eq.~(\ref{eq1}).
The relative probability for each particle to be produced is obtained by
integrating the flux spectra.
For black bodies (without grey-body factors), the relative probability
for bosons to be produced is 1 and for fermions is 3/4. 

The gravitational coupling is flavour blind and to first order a black
hole emits all 118 Standard Model particle and antiparticle degrees of
freedom with approximately equal probability.
To obtain more accurate relative rates requires knowledge of the
grey-body factors, including their full energy dependence. 
Many calculations of the grey-body factors have been performed.
In four dimensions, the relative emissivities per degree of freedom for
a non-rotating black hole are 1, 0.37, 0.11, and 0.01 for spin-0, 1/2,
1, and 2 modes. 
Kanti and March-Russell~\cite{Kanti02a,Kanti02b} calculated the grey-body
factors in higher dimensions analytically using a low-energy
approximation. 
Harris and Kanti~\cite{Harris03b,Kanti04} performed an exact calculation
of the grey-body factors numerically. 
Ida, Oda, and Park~\cite{Ida06,Ida05a,Ida02} have performed the
calculation for rotating black holes. 
The rotating case has also been performed by sets of different
authors~\cite{Casals06,Duffy05,Casals05}. 

Gravitons can be handled by considering weak perturbations from external
fields.
The perturbations are divided into scalar, vector, and tensor.
Tensor perturbations exist only in greater than four dimensions.
The total absorption cross section is obtained by summing the absorption
coefficients $\mathcal{A}_l^{(s)}(\omega)$ for each  mode $l$ weighted
by the multiplicity factors $\mathcal{N}_{n,l}^{(s)}$.   
For $n=0$, $\mathcal{N}_{0,l}^{(S)} = \mathcal{N}_{0,l}^{(V)} = (2l+1)$ and
$\mathcal{N}_{0,l}^{(T)}=0$.  
The total flux for gravitational waves is

\begin{equation} \label{eq19}
\frac{dN}{dt} = \sum_{l=2}^\infty \int \frac{d\omega}{2\pi}
\frac{1}{\exp(\omega/T_\mathrm{H}) -1} \left[ 
  \mathcal{N}_{n,l}^{(S)} \left| \mathcal{A}_{l}^{(S)}(\omega) \right|^2 
+ \mathcal{N}_{n,l}^{(V)} \left| \mathcal{A}_{l}^{(V)}(\omega) \right|^2 
+ \mathcal{N}_{n,l}^{(T)} \left| \mathcal{A}_{l}^{(T)}(\omega) \right|^2 
\right]\, ,
\end{equation}

\noindent
where the counting of helicities is included in the multiplicity
factors.

Again, knowledge of the grey-body factors is essential.
Creek, Efthimiou, Kanti, Tamvakis~\cite{Creek06} calculated the
grey-body factors for gravitons in the bulk using an analytical
approximation.
Park~\cite{Park05} performed the calculation for gravitons on the brane.
Cardoso, Cavagil{\'a},
Gualtieri~\cite{Cavaglia03c,Cardoso06b,Cardoso06a} solved for the exact
grey-body factors for gravitons in the bulk numerically.  

Table~\ref{emmis} shows the fractional emission rates per degree of
freedom normalized to the scalar field.
The results for Standard Model particles are taken from
Ref.~\cite{Harris03b} while the results for gravitons are from
Ref.~\cite{Cardoso06a}.   
The emission rates for gravitons in higher dimensions are large, but the 
graviton results includes all the helicity states and count as one
degree of freedom.  
We see that the emissivities for high dimensions are approximately those
of a black-body (BB) spectrum, except in the case of the graviton.

\TABULAR{|l|cccccccc|c|}
{\hline
$D$ & 4 & 5 & 6 & 7 & 8 & 9 & 10 & 11 & BB \\ \hline
Higgs        & 1.00 & 1.00 & 1.00 & 1.00 & 1.00 & 1.00 & 1.00 & 1.00 & 1
\\ 
fermions     & 0.37 & 0.70 & 0.77 & 0.78 & 0.76 & 0.74 & 0.73 & 0.71 & 0.75 
\\
gauge bosons & 0.11 & 0.45 & 0.69 & 0.83 & 0.91 & 0.96 & 0.99 & 1.01 & 1
\\
graviton     & 0.02 & 0.20 & 0.60 & 0.91 & 1.90 & 2.50 & 5.10 & 7.60 & 1
\\
\hline}
{Fractional emission rates per degree of freedom normalized to the
scalar field~\cite{Harris03b,Cardoso06a}.  
The graviton results include all the helicity states and count as one
degree of freedom.  
\label{emmis}}

The probabilities of emission for different particle types are given by  

\begin{equation} \label{eq20}
P_i = \frac{\epsilon_i \times \mathrm{dof}_i}{\sum_j \epsilon_j \times
\mathrm{dof}_j }\ , 
\end{equation}

\noindent
where $\epsilon_i$ and $\mathrm{dof}_i$ are the emissivity and number of 
degrees of freedom of particle $i$.
Table~\ref{probpart} shows the probabilities for different particles
types.   
We now see that graviton emission is significant but not dominant. 
Significant jets (quarks and gluons), very few photons, and insignificant
Higgs bosons should be observed. 
Using Table~\ref{probpart}, we can estimate the types of signatures in a
detector: 74\% hadronic energy, 9\% missing energy, 8\% electroweak
bosons, 6\% charged leptons, 2\% photons, and 1\% Higgs bosons.
The Standard Model particle results are consistent with earlier
results~\cite{Landsberg02p}. 
We conclude that jets will dominate black hole events while missing
energy will be significant.

\TABULAR{|l|cccccccc|c|}
{\hline
$D$ & 4 & 5 & 6 & 7 & 8 & 9 & 10 & 11 & BB\\ \hline
quarks          
& 0.71 & 0.66 & 0.62 & 0.59 & 0.57 & 0.55 & 0.53 & 0.51 & 0.56\\
charged leptons 
& 0.12 & 0.11 & 0.10 & 0.10 & 0.10 & 0.09 & 0.09 & 0.09 & 0.09\\
neutrinos       
& 0.06 & 0.06 & 0.05 & 0.05 & 0.05 & 0.05 & 0.04 & 0.04 & 0.05\\
gluons          
& 0.05 & 0.09 & 0.12 & 0.14 & 0.15 & 0.16 & 0.16 & 0.16 & 0.17\\
photon          
& 0.01 & 0.01 & 0.02 & 0.02 & 0.02 & 0.02 & 0.02 & 0.02 & 0.02\\
EW bosons       
& 0.03 & 0.05 & 0.07 & 0.08 & 0.09 & 0.09 & 0.09 & 0.09 & 0.09\\
Higgs           
& 0.03 & 0.01 & 0.01 & 0.01 & 0.01 & 0.01 & 0.01 & 0.01 & 0.01\\
graviton        
& 0.00 & 0.00 & 0.01 & 0.01 & 0.02 & 0.03 & 0.05 & 0.08 & 0.01\\
\hline}
{Probability of emission for different particles.
\label{probpart}}

\section{Binding of the black hole to the brane\label{sec4}}

Normally a black hole will not move into the bulk because it is likely
to have charge, colour, or lepton/baryon number hair that will keep it
on the brane.  
However, the emission of higher-dimensional gravitons will cause the
black hole to recoil into the extra dimensions if there is no symmetry
that suppresses the recoil.

Most studies of low-scale gravity models in large extra dimensions
assume the so called probe-brane approximation.
In this approximation, the only effect of the brane field is to bind the  
black hole to the brane, and that otherwise the black hole may be
treated as an isolated object in the extra dimensions. 
The brane must intersect the black hole orthogonally~\cite{Emparan00}.
To talk about a black hole escaping from the brane into the
higher-dimensional bulk requires us to go beyond the probe-brane
approximation.   
  
There are two main approaches used to study the escape of a black hole  
from the brane.
One is to model the domain wall as a field-theoretical topological
defect. 
The phenomena of escape is thus studied by treating the brane as a
domain wall in a scalar effective field theory.
Another approach is to treating the brane in the Dirac-Nambu-Goto
approximation, and analyze the problem by studying the interaction
between a Dirac-Nambu-Goto brane and a black hole assuming adiabatic
(quasi-static) evolution.

The recoil of a black hole was studied by Frolov and
Stojkovi{\'c}~\cite{Frolov04,Frolov02,Frolov02p} within a toy model  
consisting of two scalar fields, one describing the black hole and the
other a possible quanta emitted by the black hole in the process of
evaporation.  
The interaction with the brane was approximated as weak, and it was
shown that as soon as a quanta was emitted in the extra dimensions, the 
black hole left the brane. 
In their study, it is not clear how the separation process occurs. 
Flachi \textit{et al.}~\cite{Flachi06a} examined the problem further by
studying the interaction of a small black hole and a domain wall
composed of a scalar field, and simulated the evolution of the system
when the black hole acquires an initial recoil velocity. 

Flachi and Tanaka~\cite{Flachi05} studied the dynamics of
Dirac-Nambu-Goto branes in black hole spacetimes and suggested a
mechanism for the escape of the black hole based on reconnection of the
brane.
Once the black hole acquires an initial recoil velocity perpendicularly
to the brane, an instability develops and the brane tends to envelop the
black hole.
These results were obtained in the approximation that the tension of the
brane has no self-gravity effect.
While ignoring the tension is reasonable when the recoil velocity is
large, it might not be so in the opposite case of small recoil
velocity. 
The configuration with a black hole on a brane is stable under a
perturbation causing a small recoil velocity.

When the tension of the brane is large, the deformation of the geometry
caused by the gravity of the brane needs to be taken into account. 
It is not clear if the brane tension will prevent the black hole from
escaping for small recoil velocities.
Flachi \textit{et al.} restricted their considerations to effects which are
lowest order in the brane tension next to the probe-brane
approximation. 
A critical escape velocity was found and thus it is possible the black
hole could leave the brane before evaporation is complete if the initial
mass of the black hole is large. 
Even if the black hole leaves the brane, it feels a restoring force due
to the brane tensions and is not likely to move very far.

The height of the energy barrier for escape is approximately~\cite{Flachi06b} 

\begin{equation} \label{eq21}
\mathcal{O}\left( \sigma(G_D M)^{3/(D-3)}\right)\, ,
\end{equation}

\noindent
where $M$ is the black hole mass, $G_D$ is the $D$-dimensional Newton
constant, and $\sigma$ is the brane tension.  
Using the Dimopoulos and Landsberg definition of $G_D =
1/M_\mathrm{P}^{D-2}$, where $M_\mathrm{P}$ is the fundamental Planck
scale ($\sim$ TeV), we can write the barrier energy as    

\begin{equation} \label{eq22}
V = \sigma \left( \frac{1}{M_\mathrm{P}} \right)^3 \left(
\frac{M}{M_\mathrm{P}} \right)^\frac{3}{D-3}\, .
\end{equation}

\noindent
Typically we expect $\sigma \sim M_\mathrm{P}/l^3$, where the length is
$l\sim 1/M_\mathrm{P}$.
This gives $\sigma \sim M_\mathrm{P}^4$.
Defining the dimensionless tension

\begin{equation} \label{eq23}
\hat{\sigma} = \left( \frac{1}{M_\mathrm{P}^4} \right) \sigma\, ,
\end{equation}

\noindent
we write

\begin{equation} \label{eq24}
V = \hat{\sigma} M_\mathrm{P} \left( \frac{M}{M_\mathrm{P}}
\right)^\frac{3}{D-3}\, ,  
\end{equation}

\noindent
where $\hat{\sigma}$ is of order 1.
To leave the brane, a black hole must have a momentum $p_\perp$
transverse to the brane given by

\begin{equation} \label{eq25}
p_\perp > \sqrt{V(V+2M)}\, .
\end{equation}

We ignore the rare possibility of a black hole reentering the brane
after escaping to the bulk.
Such scenarios have been examined by Dvali \textit{et al.}~\cite{Dvali01b}.

\section{Black hole recoil model and simulation\label{sec5}}

To study the effects of missing energy in black hole decay, we have
constructed a simple model.
All Standard Model particles evaporating from the black hole do so in
four dimensions. 
The graviton evaporates off the black hole in $D$ dimensions.
We assume the evaporated particles are non-interacting so that the
graviton is free to move in the extra dimensions without impediment just
like the Standard model particles move in four dimensions. 
We also assume the graviton is massless and does not decay or otherwise
interact in the bulk.
This approximation is justified when the tension of the brane is small
so that the interaction between the Nambu-Goldstone boundary fields and
the KK modes is exponentially suppressed. 
Soft branes also reduce graviton interactions with Standard Model
particles~\cite{Murayama}. 

In our model, the black hole is treated differently because of its mass
and is bound to the brane by a brane tension according to the model
of Ref.~\cite{Flachi06a,Flachi05,Flachi06b}.  
We perform the decay kinematics in $D$ dimensions and keep track of the
black hole recoil momentum transverse to the brane.
If this momentum exceeds the binding potential of the black hole to the
brane (eq.~(\ref{eq25})), the black hole is considered to have escaped
from the brane; the decay process is stopped and the missing energy is
recorded.  

Details of the horizon formation, balding, and spin-down phases have
been ignored. 
The important effects of angular momentum in the production and decay of
the black hole in extra dimensions are not taken into accounted. 
Our black holes can be considered as $D$-dimensional Schwarzschild
solutions.

We implemented the Hawking evaporation phase in two steps: determination
of the particle types and assigning energy to the decay products. 
Particle types are randomly selected with a probability determined by
their number of degrees of freedom and the ratio of emissivities.
To pick between a particle or antiparticle, the emitted charge and
baryon number are chosen such that the magnitude of the black hole
charge and baryon number does not increase after a particle is emitted,
else particles and antiparticles are chosen with equal probability.
All Standard Model particles are considered included a
Higgs\footnote{Include the scalar Higgs is not significant since it has
only one degree of freedom in all dimensions.}.

The energy assigned to the decay particles in the evaporation phase has
been implemented as follows. 
The particle type selected as described by the model above is given a 
random energy according to its decay spectrum. 
A different decay spectrum is used for scalars, fermions, and vector
bosons, i.e.\ the spin statistics factor is taken into account. 
Grey-body factors for Standard Model particles are used without
approximations~\cite{Harris03b}.   
Spectra for massless particles are used, even for the gauge bosons and
heavy quarks. 
This is a good approximation provided the top-quark mass $m_\mathrm{t}
\ll T_\mathrm{H}$. 
The Hawking temperature is updated after each particle is emitted.
This assumes the decay is quasi-stationary in the sense that the black
hole has time to come to equilibrium at each new temperature before
the next particle is emitted.
The energy of the particle given by the spectrum must be constraint to
conserve energy and momentum at each step.

The evaporation phase ends when the black hole mass drops below the
Planck scale.
When this occurs, a final isotropic two-body phase-space decay is
performed.  
The black hole decays totally to Standard Model particles and/or
gravitons. 
Overall electric charge, baryon number, and colour are conserved in the 
black hole production and decay.

If the black hole escapes from the brane during evaporation, up to two
partons with the black hole charge and baryon number are added to the
event record with zero momentum.  
In this way, we can complete the colour connection, yet still account
for missing energy.

To implement our model, we started from the Monte Carlo event generator
CHARYBDIS version 1.003~\cite{Harris03a,Gingrich06b} and adapted it for
our study.  
It was interfaced to PYTHIA which provides the parton evolution and
hadronization, as well as Standard Model particle decays. 
The interface to PYTHIA or HERWIG is not important since the studies
presented here are at the particle level. 
Gravitons were added as a particle type and the kinematics for the
evaporation of the graviton from the black hole were calculated in $D$
dimensions. 
The condition for escape from the brane was examined after each
graviton was emitted.
Table~\ref{charybdis} lists the CHARYBDIS parameter settings.

\TABULAR{|c|l|c|}
{\hline
Name   & Description                                & Value   \\ \hline
MINMSS & Minimum mass of black holes                &  5 GeV  \\
MAXMSS & Maximum mass of black holes                & 14 GeV  \\
MPLNCK & Planck scale                               &  1 GeV  \\
MSSDEF & Planck scale definition                    &  2      \\
TOTDIM & Total number of dimensions                 & 6, 8, or 11 \\
NBODY  & Number of particles in remnant decay       & 2       \\ 
GTSCA  & Black hole mass used as PDF momentum scale & true    \\
TIMVAR & Allow $T_\mathrm{H}$ to change with time   & true    \\
MSSDEC & Use all SM particles as decay products     & true    \\
GRYBDY & Include grey-body effects                  & true    \\
KINCUT & Use a kinematic cut-off on the decay       & false   \\
\hline}
{Parameters use in the CHARYBDIS generator.
\label{charybdis}}

The graviton is represented by a zero charged, non-interacting, massless
particle in $D$ dimensions. 
The $(D-4)$ extra dimensions are represented internally in CHARYBDIS and 
are not known to PYTHIA or appear in the event record.
The black hole is also treated as a $D$-dimensional particle internally
in  CHARYBDIS.
The black hole is not added to the event record since it decays
entirely in CHARYBDIS.
See Koch, Bleicher, Hossenfelder~\cite{Koch05} for an alternative
formulation. 

The charge and baryon number of the black hole are recalculated after
each particle is emitted.
The final two-body decay must generate two particle that have the charge
and baryon number of the black hole.
Sometimes the black hole will have too high an absolute charge or baryon
number so that a two-body final state is not possible.
We thus encourage the absolute charge and baryon number of the black
hole not to become too large by choosing between particle or
antiparticle states to minimize the absolute charge and baryon number of
the black hole after each decay.
For example, a quark or anti-quark will be chosen to reduced the
absolute value of the black hole baryon number after the decay.
A similar difficulty can occur when the black hole leaves the brane with
a large charge or baryon number.
We need to include a number of zero-energy quarks or anti-quarks in the
event to allow the colour connection to be made.

If the black hole does not leave the brane, a final two-body decay is
performed. 
Three possibilities exist for the final decay:
1) both particles are Standard Model particles, CHARYBDIS performs a
normal decay in four dimensions, 2) one is a Standard Model particle and
one is a graviton, CHARYBDIS performs the decay in four dimensions, and
3) both particles are gravitons, a two-body decay is performed in $D$
dimensions.     
The special case 2) of the graviton being restricted to four dimensions
is not important since there is no longer a black hole to recoil against
it.  
In any case, gravitons will appear as missing energy on the brane.

\section{Results\label{sec6}}

We present results for black holes with $5<M<14$~TeV and
$M_\mathrm{P}=1$~TeV. 
Since the cross section falls steeply with increasing black hole
mass, most of the black holes will have a mass close to 5~TeV while
very few will have a mass above 9~TeV. 
Most results will be presented in 11 dimensions but sometimes six and
eight dimensions will be used for comparison.
Normally we examine the results under two extreme choices of brane
tension: vanishing tension $\hat{\sigma}=0$ and strong tension
$\hat{\sigma}=10^3$.   

\subsection{Graviton\label{sec6.1}}

We studied the effects of adding the graviton to CHARYBDIS by examining the
distribution of particle types from black hole decay.   
CHARYBDIS relies on conserving overall charge and baryon number in the
final decay. 
Differences from the probabilities in Table~\ref{probpart} can be due to
requiring charge and baryon number conservation, as well as
energy-momentum conservation.
To study these asymmetries, we first generated $\mathrm{p\bar{p}}$
collisions with black-body spectra for the emitted particles.   
We eliminated the need to conserve overall charge, baryon number, and
energy-momentum in the Hawking evaporation.  
The fractional occurrences of the different particle types were as
expected (Table~\ref{probpart}) to an accuracy better than 0.8\%.  

Having simulated the relative frequency of occurrence for different
particles in Hawking evaporation according to expectations, we simulated
pp collisions and included grey-body factors, as well as the two-body
final decay.   
The resulting frequency of occurrence of each particle type is shown in
fig.~\ref{pid}.
The simulated results are shown as the black histogram and the
expectations, according to black-body spectra in four dimensions, as the
red histogram. 
We see that quarks are enhanced over anti-quarks and gluons in order to
conserve the normally positive baryon number of the initial state.
Similarly, positive charged quarks and anti-quarks are enhanced over
negative charged particles in order to conserve the normally positive
net charge of the black hole. 
Differences between the simulation and black-body democracy can be as
high as 80\%. 
The graviton frequency of occurrence is more than 660\% times higher than
that predicted by the  black-body spectrum in four dimensions.
Besides the asymmetries due to grey-body factors, asymmetries occur
during the two-body final decay.  
If there is a need to conserve other quantum numbers, like lepton
number, further asymmetries will result. 

\EPSFIGURE{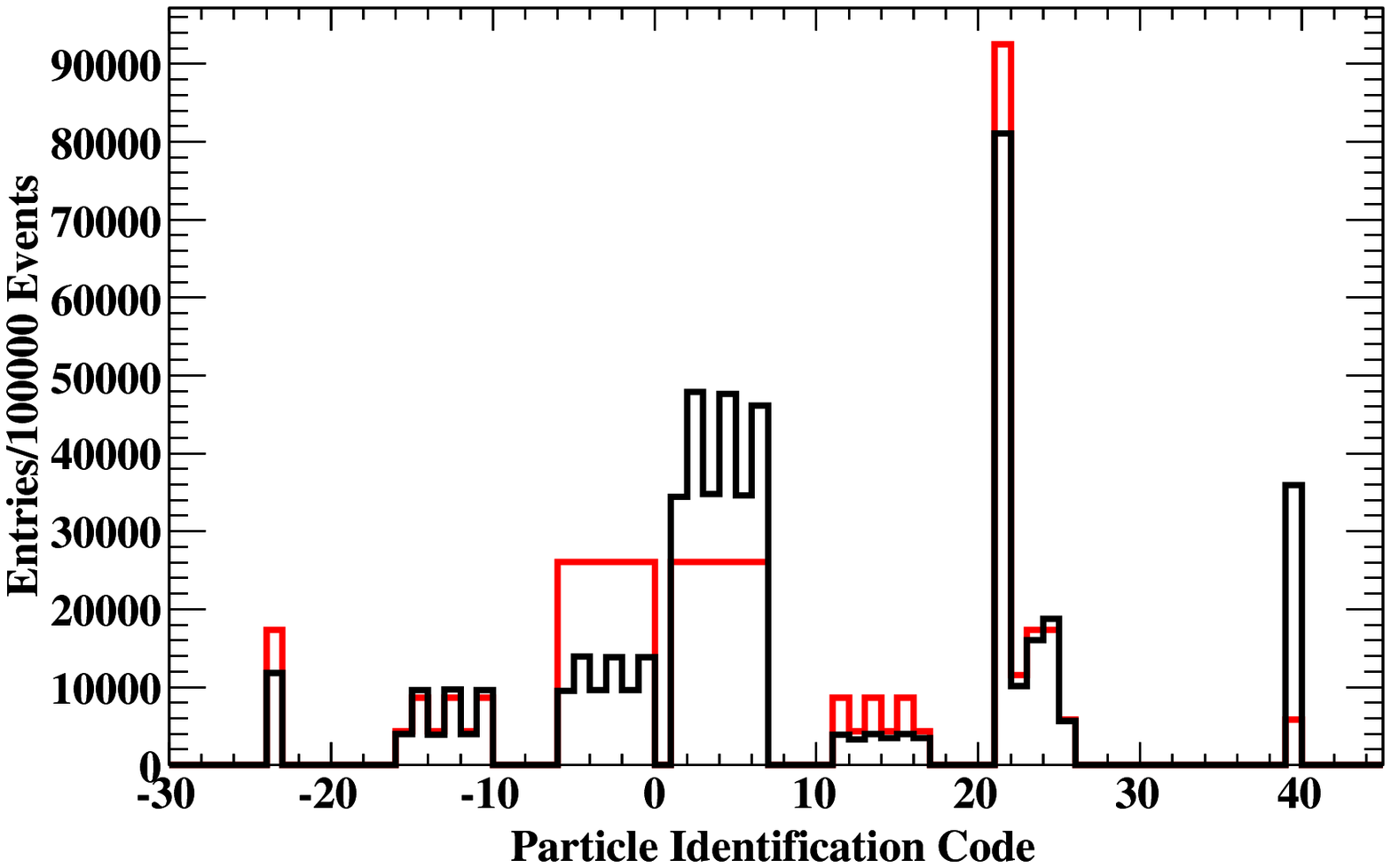,width=12cm}
{Frequency of occurrence of each particle type (particle identification
code) for black holes with $5<m<14$~TeV, $M_\mathrm{P}=1$~TeV, and
$D=11$. 
The simulated results are shown as the black histogram while the
black-body expectations are shown as the red histogram.
The positive particle identification codes are 1 d-quark, 2 u-quark, 3
s-quark, 4 c-quark, 5 b-quark, 6 t-quark, 11 e$^-$, 12 $\nu_\mathrm{e}$,
13 $\mu^-$, 14 $\nu_\mu$, 15 $\tau^-$, 16 $\nu_\tau$, 21 gluon, 22 photon,
23 Z, 24 W$^+$, 25 Higgs, 39 graviton. 
The negative particle identification codes are the antiparticles.
\label{pid}}

The emissivity of gravitons in $D$ dimensions has been
calculated~\cite{Cardoso06b,Cardoso06a}.
What is not known, or readily available, is the shape of the decay
spectrum for gravitons.
The Standard Model flux spectra~\cite{Harris03b} are shown in
fig.~\ref{norm}.  
Each spectrum has been normalized to unit area.
We see that the relative shapes of the spectra are similar except for
the grey-body spin-1 case.
They are most similar in seven extra ($D=11$) dimensions and the
grey-body spin-1/2 case is a typical spectrum.
The shape of the grey-body spectrum for spin-1/2 particles was used for
the graviton.
The sensitivity to this arbitrary choice is examined in
section~\ref{sec6.4}.  

\EPSFIGURE{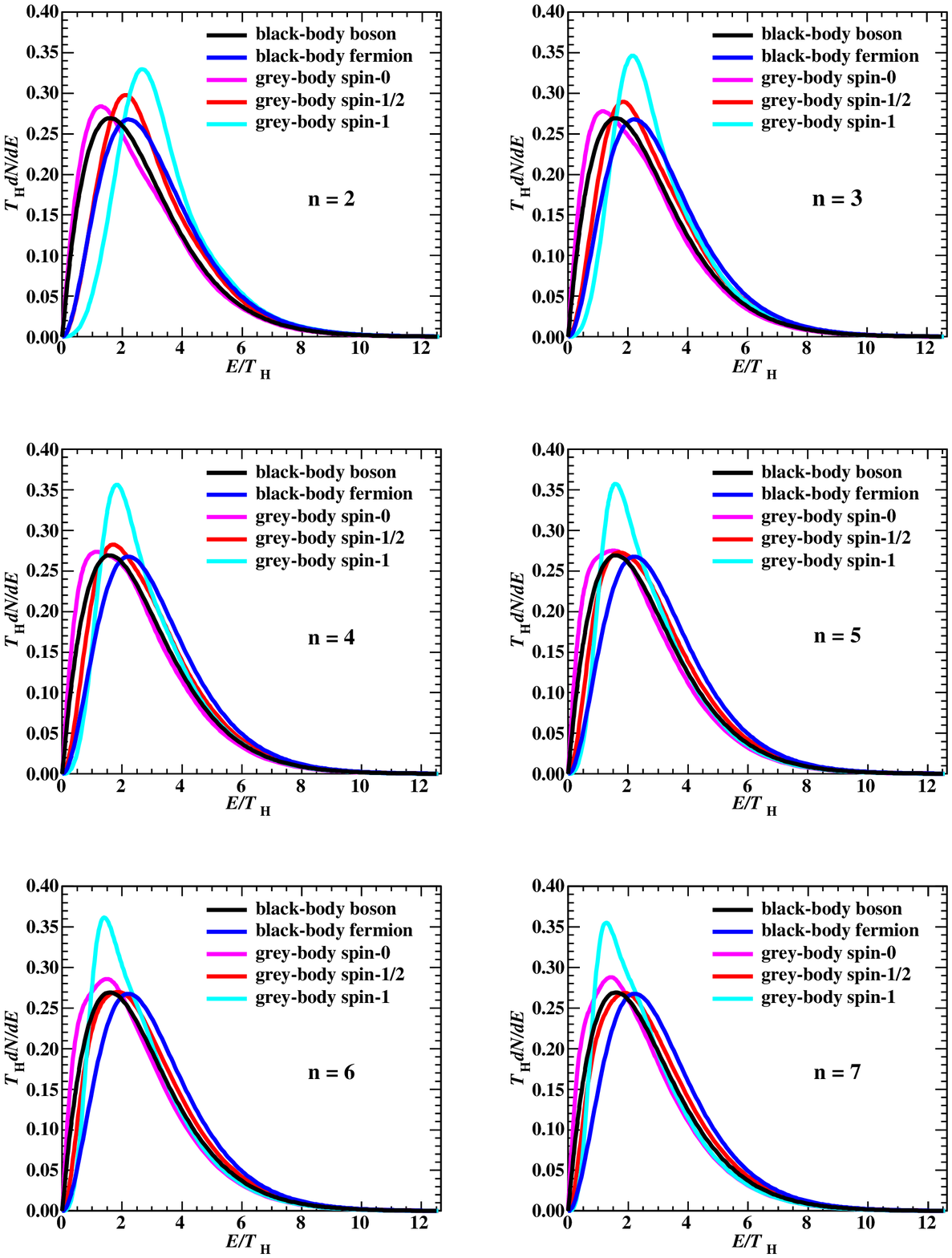,width=14cm}
{Energy spectra for grey bodies and black bodies~\cite{Harris03b}.
Each spectrum has been normalized to unit area.
\label{norm}}

\subsection{Recoil effect\label{sec6.2}}

We can ask at what point during Hawking evaporation we would expect the
black hole to be sufficiently perturbed to leave the brane?
Assuming the kinetic energy must be greater than the barrier potential,
the critical perpendicular velocity for a black hole to leave the brane
is 

\begin{equation} \label{eq26}
v_\mathrm{c} = \sqrt{\hat{\sigma}} \left( \frac{M_\mathrm{p}}{M}
\right)^\frac{D-6}{2(D-3)}\, .  
\end{equation}

\noindent
The average black hole recoil velocity after evaporating off a particle
is of the order~\cite{Flachi06b}

\begin{equation} \label{eq27}
v_\mathrm{r} = \left( \frac{M_\mathrm{p}}{M}
\right)^\frac{D-2}{2(D-3)}\, .  
\end{equation}

\noindent
Thus the black hole would leave the brane when $v_\mathrm{r} \gtrsim
v_\mathrm{c}$ which should occur at the critical mass

\begin{equation} \label{eq28}
M \lesssim M_\mathrm{c} =\frac{M_\mathrm{P}}{\hat{\sigma}^{(D-3)/4}}\, .
\end{equation}

\noindent
If the initial mass of the black hole is greater than $M_\mathrm{c}$, it
may decay down to $M_\mathrm{c}$. 
If the initial mass is below $M_\mathrm{c}$, the black hole will
probably leave the brane when the first graviton is emitted. 
If the black hole minimum mass is above $M_\mathrm{c}$, the black hole
will not leave the brane. 
In our model, the minimal black hole mass is $M_\mathrm{P}$, so the black
hole will only leave the brane if $\hat{\sigma} \lesssim 1$.

Figure~\ref{brane} shows the probability per event for a black hole with
$5 < M < 14$~TeV and $M_\mathrm{P}=1$~TeV to leave the brane for
different brane tensions $\hat{\sigma}$. 
The tension must be weak $(\hat{\sigma}<1)$ for the black hole to have a
significant probability to leave the brane.   
In the extreme case of very weak tension ($\hat{\sigma} \to 0$), the
probability becomes 6.9\% for $D=6$ and 33.4\% for $D=11$.
In this case, the black hole will normally leave the brane as soon as
the first graviton is emitted.
We would expect similar shaped curves to fig.~\ref{brane} for different
dimensions, Planck scales, and black hole masses.
We might expect the probability to leave the brane at zero tension to
increase in lower dimensions because of the increase in particle
multiplicity.
However, the probability of graviton emission per evaporated particle
decreases with lower dimensions, so that the resulting probability per
event for the black hole to leave the brane is lower in lower
dimensions. 
Thus 33\% is likely the maximum probability for black holes to leave
the brane at LHC energies for $D\le 11$.

\EPSFIGURE{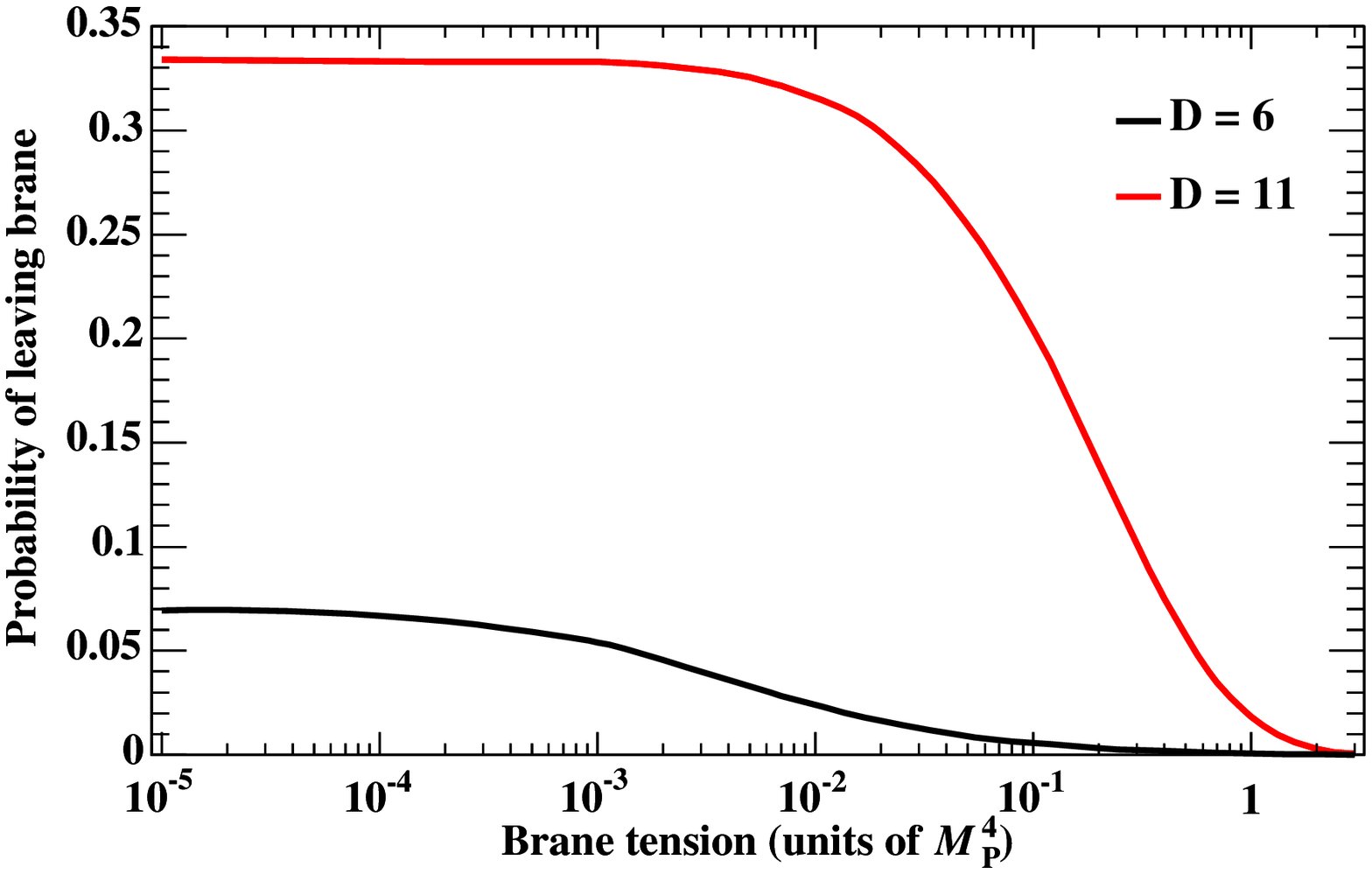,width=12cm}
{Probability per event that a black hole will leave the brane versus
 brane tension for $5<M<14$~TeV and $M_\mathrm{P}=1$~TeV. 
The black curve is for $D=6$ while the red curve is for $D=11$.
\label{brane}}

\subsection{Multiplicities\label{sec6.3}}

Multiplicities and emission probabilities for different particles, in
particular the graviton, will be different for Hawking evaporation, the
two-body final decay, and for events in which the black hole leaves the
brane. 
Figure~\ref{mult} show multiplicity distributions of primary particles
emitted from black holes with $5<M<14$~TeV and $M_\mathrm{P} = 1$~TeV
for $D=6$ (fig.~\ref{mult}a) and $D=11$ (fig.~\ref{mult}b). 
These distributions include particles emitted by Hawking evaporation, as
well as the two particles from the final decay.
The black histograms are for all primary particles, and have means of
9.3 and 5.6 for $D=6$ and $D=11$.
The red histograms in fig.~\ref{mult} are for the case of only visible
primary particles when the black hole is allowed to leave the brane with
$\hat{\sigma}=0$. 
In this case, the mean multiplicities drop to 8.7 and 4.8 for $D=6$ and
$D=11$, where the bin for zero multiplicity has not been included in the   
calculation of the means.

\EPSFIGURE{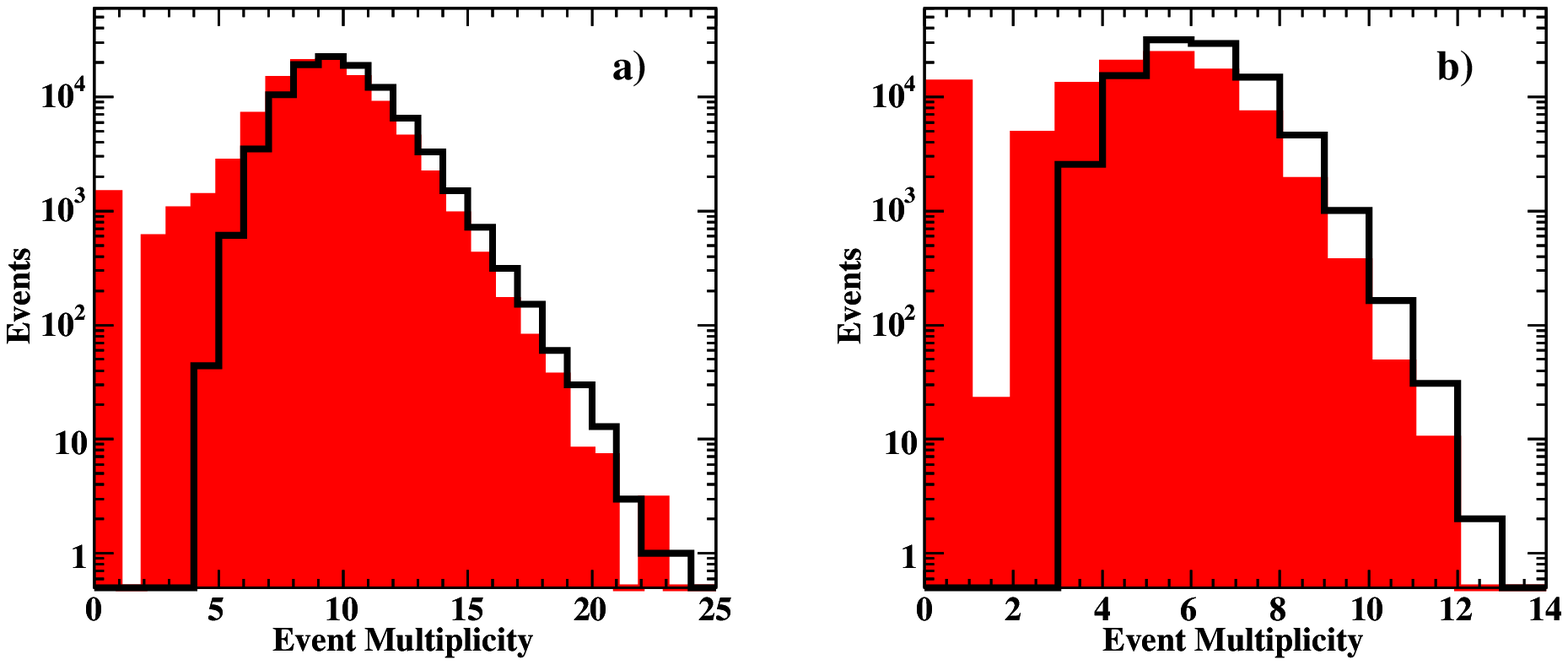,width=15cm}
{Multiplicity distributions of primary particles emitted from black
holes with $5<M<14$~TeV and $M_\mathrm{P}=1$~TeV for a) $D=6$ and b)
$D=11$.  
The black distributions are for all primary particles while the red
distributions are for visible primary particles only and the black hole
is allowed to leave the brane before the evaporation process is
completed.   
\label{mult}}

From the zero bins in fig.~\ref{mult}, we seen that 1.4\% and 13.3\% of
the events will have no visible particles for $D=6$ and $D=11$. 
Most of these events correspond to the extreme case when the black hole
emits a graviton as the first particle and immediately leaves the brane,
with probabilities 1.4\% and 12.8\% for $D=6$ and $D=11$.
It is also possible with probabilities 0.07\% and 0.6\% for $D=6$ and
$D=11$ that the black hole first emits neutrinos and then a graviton,
and leaves the brane.   
For these two cases, there would be no evidence that the black hole was  
ever formed or a proton-proton collision occurred.  
We would under-measure the black hole production rate and cross
section by 1\% and 13\% for $D=6$ and $D=11$ with no possibility to
correct the measurement based on the data itself.

Based on the average multiplicities and probability of graviton
emission, we can estimate the asymptotic values in fig.~\ref{brane}.  
For example, since the probability of emitting a graviton per emission
is 12.8\% and the mean multiplicity is 4.8 in 11 dimensions for
$\hat{\sigma}= 0$, we estimate the probability per event to leave the
brane is 35\% which compares favourably with the simulated result of
33\%.  

Table~\ref{grav} shows the percentage of gravitons produced in all
events for strong tension $\hat{\sigma}=10^3$ and zero tension. 
Multiple graviton emission per event in 11 dimensions is significant
(4\%) and comparable to single graviton emission in six dimensions
(5\%).

\TABULAR{|c|c|c|c|c|c|c|}
{\hline
Number & \multicolumn{2}{c|}{$D=6$} & \multicolumn{2}{c|}{$D=8$} & 
\multicolumn{2}{c|}{$D=11$} \\ \cline{2-7}
Gravitons & $\hat{\sigma}=10^3$ & $\hat{\sigma}=0$ &
$\hat{\sigma}=10^3$ & $\hat{\sigma}=0$ & 
$\hat{\sigma}=10^3$ & $\hat{\sigma}=0$ \\ \hline
0 & 94.8 & 92.2 & 89.3 & 85.9 & 69.1 & 61.3\\
1 &  5.0 &  7.8 & 10.2 & 14.1 & 26.4 & 38.6\\
2 &  0.1 &  0.0 &  0.5 &  0.0 &  4.2 &  0.2\\
3 &  0.0 &  0.0 &  0.0 &  0.0 &  0.4 &  0.0\\
\hline}
{Percentage occurrence of various number of gravitons per event in
black hole decay.  
For $\hat{\sigma}=10^3$ the black hole does not leave the brane while
for $\hat{\sigma}=0$ the black hole may leave the brane.
\label{grav}}

For comparison, we show the corresponding numbers for neutrinos in
Table~\ref{neut}.
For $\hat{\sigma}=10^3$, more particles are emitted so the multiplicity
of neutrinos and multiple neutrino emission is higher than for the
$\hat{\sigma}=0$ case. 
For gravitons, the situation is reversed since the graviton ends the
decay and thus inhibits other particles evaporating off the black hole.   

\TABULAR{|c|c|c|c|c|c|c|}
{\hline
Number & \multicolumn{2}{c|}{$D=6$} & \multicolumn{2}{c|}{$D=8$} & 
\multicolumn{2}{c|}{$D=11$} \\ \cline{2-7}
Neutrinos & $\hat{\sigma}=10^3$ & $\hat{\sigma}=0$ &
$\hat{\sigma}=10^3$ & $\hat{\sigma}=0$ & 
$\hat{\sigma}=10^3$ & $\hat{\sigma}=0$ \\ \hline
0 & 61.5 & 62.9 & 73.7 & 75.8 & 80.0 & 84.3\\
1 & 30.1 & 29.2 & 22.9 & 21.0 & 18.2 & 14.3\\
2 &  7.2 &  6.8 &  3.1 &  2.9 &  1.7 &  1.3\\
3 &  1.0 &  1.0 &  0.3 &  0.2 &  0.1 &  0.1\\
4 &  0.1 &  0.1 &  0.0 &  0.0 &  0.0 &  0.0\\
\hline}
{Percentage occurrence of various number of neutrinos per event in
black hole decay.  
For $\hat{\sigma}=10^3$ the black hole does not leave the brane while
for $\hat{\sigma}=0$ the black hole may leave the brane.
\label{neut}}

We can understand the numbers in Table~\ref{grav} from the graviton
emission probability and the particle multiplicities.
Since the probability to leave the brane for $\hat{\sigma}=0$ is equal
to the probability for graviton emission, we can use the values in the
$D=11$ and $\hat{\sigma} = 0$ column of Table~\ref{grav} to
estimate the probabilities of 0, 1, and 2 gravitons being emitted in the
two-body final decay as 91.5\%, 8.4\%, and 0.3\%.
The values for strong brane tension in the $D=11$ and $\hat{\sigma} =
10^3$ column can be understood as follows. 
Using the results for zero graviton emission, we predict a multiplicity
of 3.2 for a graviton emission probability of 7.6\%
(Table~\ref{probpart}), or we predict a graviton emission probability of
7.5\% assuming the mean multiplicity of 3.6.  
The multiplicity of 3.6 excludes the two particles from the final decay.
The results are thus consistent with each other.
Based on the 7.6\% graviton emission probability and multiplicity of 3.6,
we predict the probabilities of emitting 0, 1, 2, and 3 gravitons as
68.8\%, 26.8\%, 4.3\%, and 0.4\%. 
These predictions are consistent with the simulation results presented
in Table~\ref{grav}. 
We might expect multiple graviton emission to increase with multiplicity
for lower dimensions, but the probability of a graviton per emission
decreases with lower dimensions, and thus multiple graviton emission
becomes even less in lower dimensions. 
Multiple graviton emission should occur at a level of less than 5\% for
any brane tension at the LHC provided $D\le 11$.

\subsection{Missing energy\label{sec6.4}}

Until now, we have been talking about missing energy which is due to
undetectable particles on the brane or gravitons in the bulk. 
In proton-proton collisions, we neither know the initial-state energy or
longitudinal momentum that went into producing the black hole. 
What we do know is that the transverse momentum to the proton beams is
zero for the initial state. 
Thus the signature of missing energy can only be inferred by a non-zero
total transverse momentum in the event.
In this section, we will be more precise and talk about missing
transverse momentum $\slashed{p}_\mathrm{T}$ rather than missing energy.  

Figure~\ref{pt} shows the missing transverse momentum distribution for
black holes with $5 < M < 14$~TeV, $M_\mathrm{P} = 1$~TeV, and $D=11$. 
The black histogram shows the case when $\slashed{p}_\mathrm{T}$
is due to the three generations of neutrinos only while the red
histogram is the case for the gravitons only. 
The blue histogram is the case when the neutrinos and gravitons
contribute to $\slashed{p}_\mathrm{T}$, but the black hole is not
allowed to leave the brane ($\hat{\sigma} = 10^3$).   
The magenta histogram is the case when the black hole is allowed to
leave the brane with a vanishing brane tension.  
Events which do not emit neutrinos or gravitons (55\%) are not
included in the histogram.   
Some events emit a graviton as the first particle, and the black hole
leaves the brane without emitting a Standard Model particle.
These events have $\slashed{p}_\mathrm{T}=0$ and are also not included
in the histogram.   
More events have missing transverse momentum due to neutrinos than
gravitons, and the value of $\slashed{p}_\mathrm{T}$ from neutrinos is
higher because all of the momentum components for the neutrino are on
the brane. 
Allowing the black hole to escape form the brane increases the mean
missing transverse momentum considerably to 960~GeV for 38\% of the
events.
The other 62\% of the events have no significant
$\slashed{p}_\mathrm{T}$. 
The missing transverse momentum decreases to 530~GeV in six dimensions
for 42\% of the events.
In this case, the missing transverse momentum is predominated due to
neutrinos.  

\EPSFIGURE{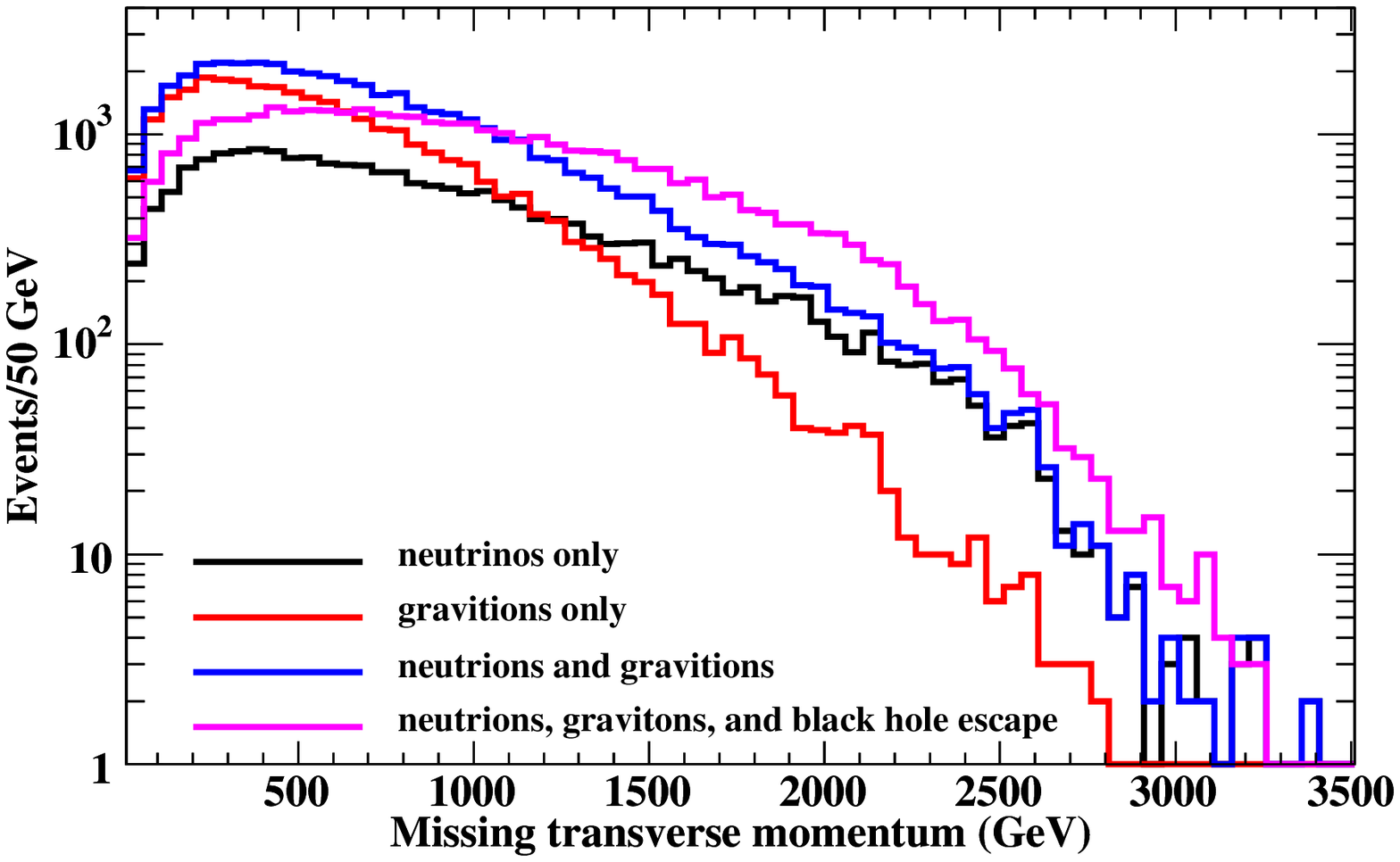,width=12cm}
{Missing transverse momentum for black holes with $5 < M < 14$~TeV,
$M_\mathrm{P} = 1$~TeV, and $D=11$. 
The black histogram is due to neutrinos only, the red histogram is due
to gravitons only, the blue histogram is due to neutrinos and gravitons
while the magenta histogram is due to neutrinos, gravitons, and the
possibility for the black hole to leave the brane.   
100,000 events are in each histogram, but events with
$\slashed{p}_\mathrm{T}<10$~GeV are zero suppressed. 
\label{pt}}

The missing transverse momentum distributions in fig.~\ref{pt} are
consistent with previous results that include the neutrinos
only and detector effects~\cite{Harris05a}. 
The black hole missing transverse momentum distributions for 11
dimensions are very different from QCD and SUSY events~\cite{Harris05a}. 

The missing transverse momentum distribution is not very sensitive to
our choice of graviton energy spectrum.
Figure~\ref{sense} shows the missing transverse momentum distribution due
to neutrinos and gravitons only for different grey-body spectra for the
graviton. 
The black histogram is for a spin-1/2 spectrum, the red histogram is for
a spin-1 spectrum while the blue histogram is for a spin-0 spectrum. 
Thus, we are insensitive to the choice of graviton spectrum provided
$\slashed{p}_\mathrm{T} \gtrsim 300$~GeV.   

\EPSFIGURE{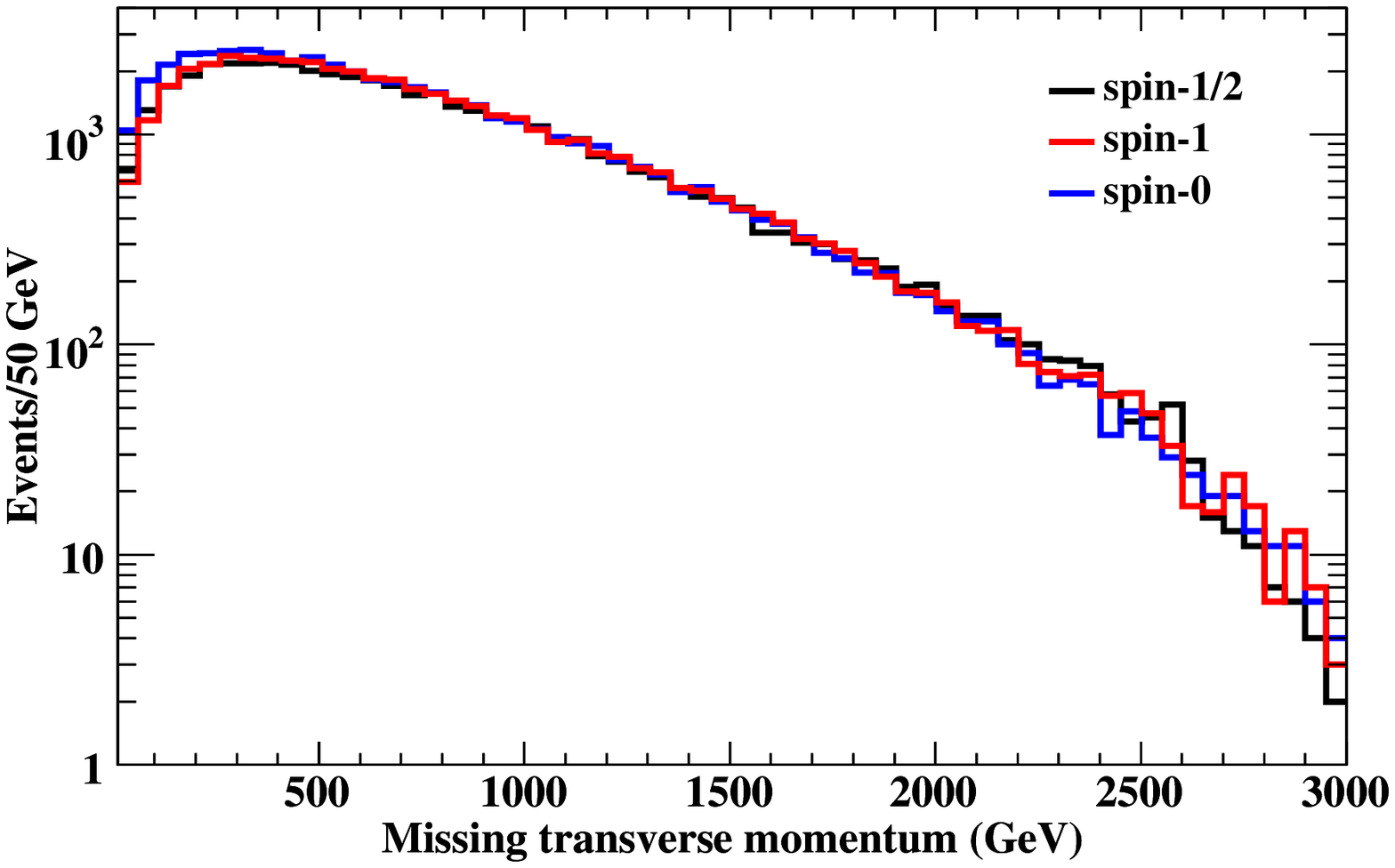,width=12cm}
{Missing transverse momentum for black holes with $5 < M < 14$~TeV,
$M_\mathrm{P} = 1$~TeV, and $D=11$ for different graviton grey-body
spectra. 
The black histogram is for a spin-1/2 spectrum, the red histogram is for
a spin-1 spectrum while the blue histogram is for a spin-0 spectrum.
100,000 events are in each histogram, but events with
$\slashed{p}_\mathrm{T}<10$~GeV are zero suppressed. 
\label{sense}}

\subsection{Mass and cross section\label{sec6.5}}

We can ask how graviton emission and black hole recoil affect
experimental measurements?
An experiment needs to first identify black hole events and then measure
the black hole mass.
We will assume the black hole events are well identified by their decay
to high-$p_\mathrm{T}$ objects and possibly missing energy.
However, events with no or little visible energy will not be identified
as black hole events, or events of any type.
Based on the multiplicity distribution, we expect a maximum of 13\% of
the black hole events to fall into this category for $D\le 11$.

Having identified the black hole events, we now need to reconstruct
their masses.
Since the black hole mass is reconstructed by summing the four-vectors
of all the particles, missing energy will result in decreasing the
reconstructed black hole masses.
However, events without neutrinos or gravitons should have well
reconstructed mass.
We expect about 60\% of the events in 11 dimensions will not be affected
by missing energy.   
Figure~\ref{scat} shows the reconstructed black hole mass versus missing
transverse momentum in 11 dimensions for vanishing brane tension.

\EPSFIGURE{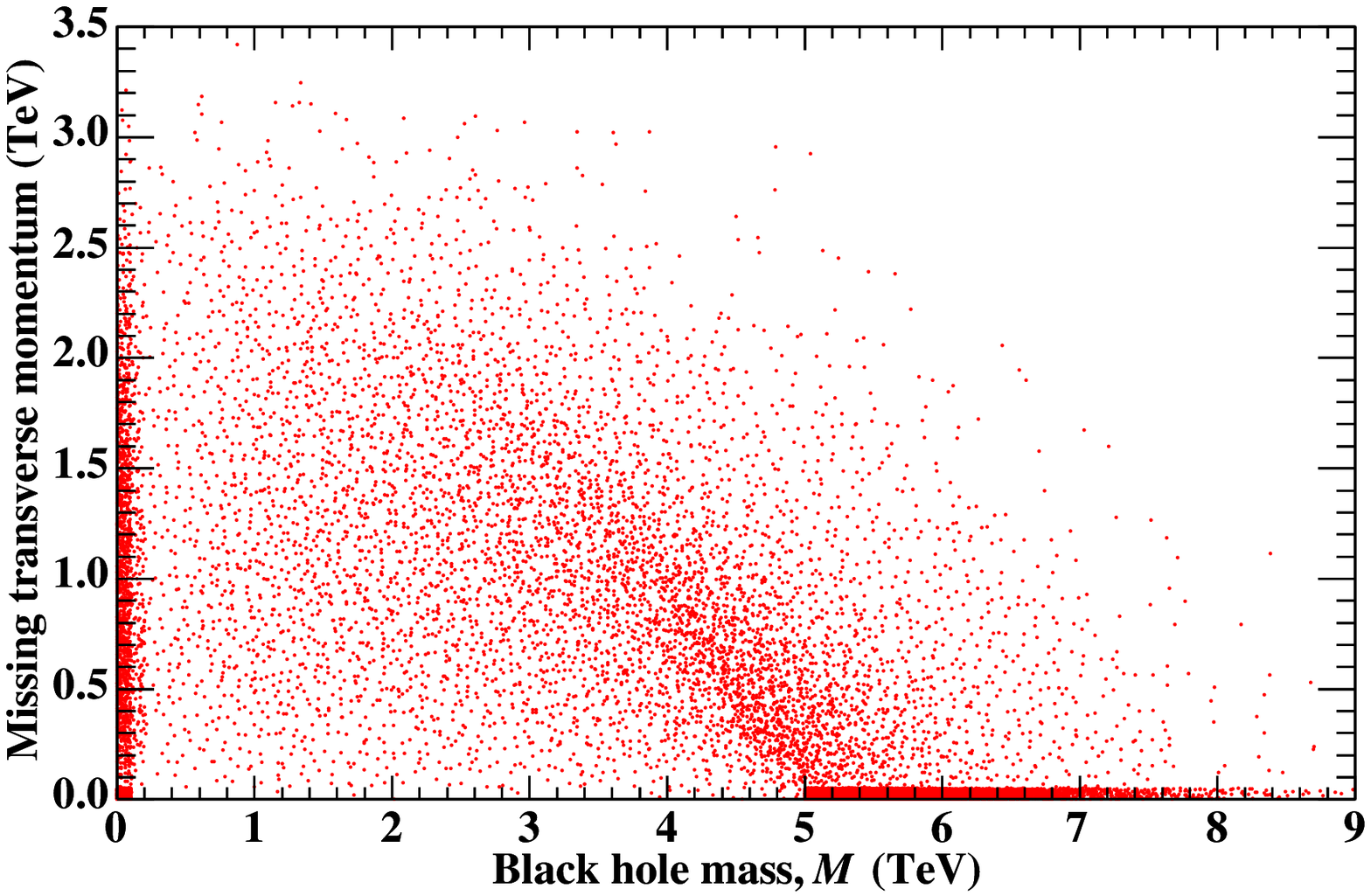,width=12cm}
{Black hole mass versus missing transverse momentum for black holes with
$5 < M < 14$~TeV, $M_\mathrm{P} = 1$~TeV, $D=11$, and vanishing brane
tension. 
\label{scat}}

In events with missing energy, the reconstructed mass will always be
low.
Black holes with mass near 5~TeV will be reconstructed with masses that 
fall outside the $5 < M < 14$~TeV mass window we are considering.
Black holes with mass well above 5~TeV will also be reconstructed with
lower masses but remain within the mass window we are considering.
Because of the steeply falling cross section with black hole mass,
the problem of migration of high-mass values to lower masses within our
mass window will be less significant than the number of events migrating
out of the mass window below 5~TeV. 
The net effect will be to decrease the total number of events
reconstructed and the shape of the differential cross section.
Selecting only events with low missing energy will decrease its effect
on the mass reconstruction and cross section determination.

As an illustrative example, we have required $\slashed{p}_\mathrm{T} <
10$~GeV and plotted the differential cross section versus black hole
mass with and without missing energy as shown in fig.~\ref{mass}. 
The exact value for the missing energy cut will have to be determined
from a full simulation of the detector and the data.
The shape of the cross section changes only slightly at high masses,
where there are few events.
The contamination of any mass bin due the migration of higher-mass
events was determined from the simulation to be of the order of 0.01\%. 

\EPSFIGURE{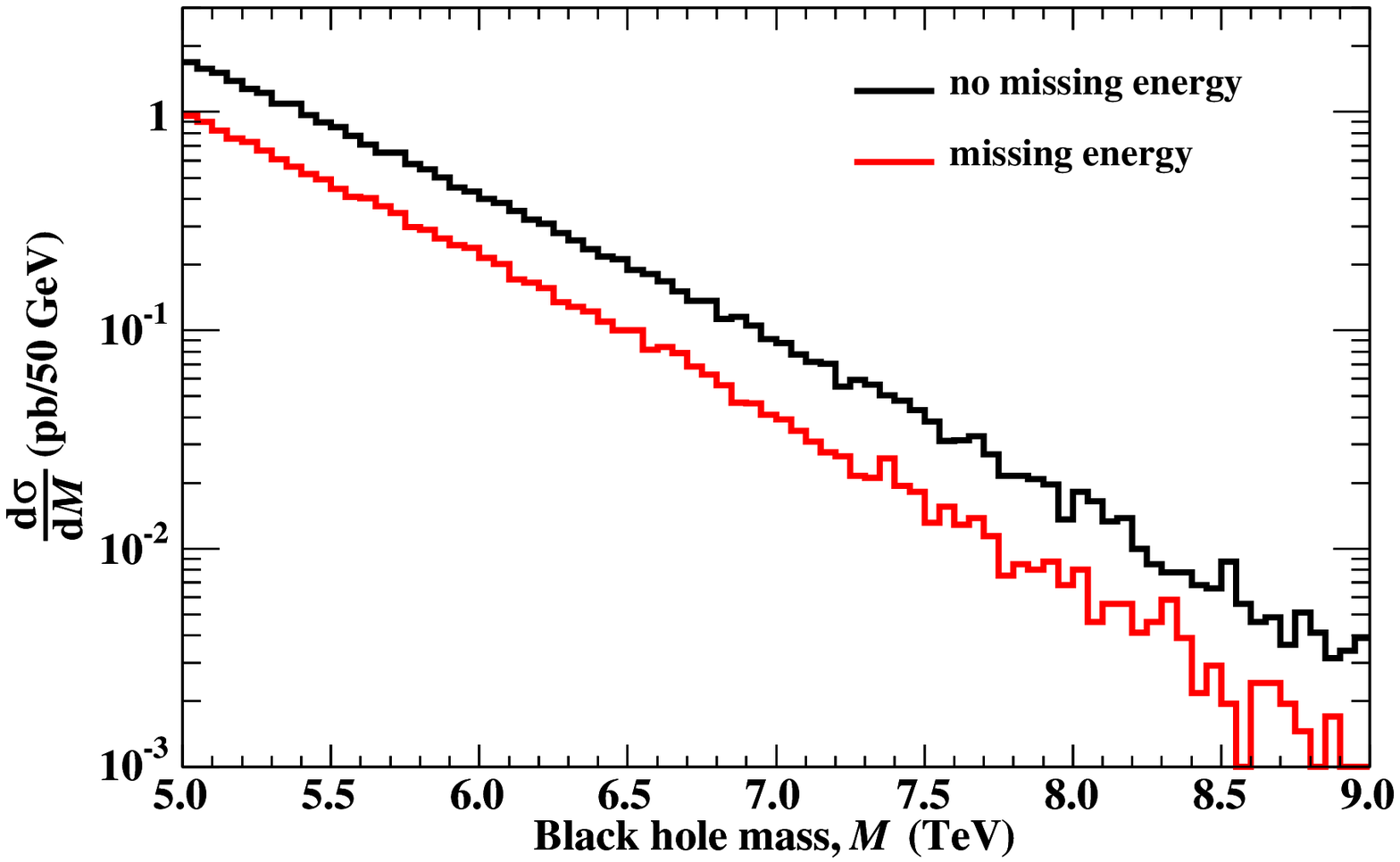,width=12cm}
{Differential cross section for black holes with $5 < M < 14$~TeV,
$M_\mathrm{P} = 1$~TeV, and $D=11$. 
The black histogram is the perfect situation in which we can
unrealistically determine the exact black hole mass.
The red histogram is due to undetected neutrinos, gravitons, and the 
possibility for the black hole to leave the brane.   
\label{mass}}

The cross section for black hole production with $5 < M < 14$~TeV,
$M_\mathrm{P}=1$~TeV, and $D=11$ is 24~pb, for the default parton
density functions used by PYTHIA. 
Integrating the differential cross section for missing energy events
(red histogram) in fig.~\ref{mass} gives a reduced cross section of
13~pb (12~pb before acceptance correction). 
Although this is only a reduction in the theoretical cross section of
about 50\%, such an error could make the determination of the Planck
scale and number of dimensions difficult. 
It may be possible to improve the mass resolution by treating the
missing transverse momentum as a massless pseudo-particle in the
calculation of the black hole mass.  
To go further in these studies, will require including particle
fragmentation, hadronization, decay, and detector effects. 
For example, the limit geometrical acceptance of detectors and
heavy-particle decays will further contribute to the missing energy.  

\section{Discussion\label{sec7}}

We now discuss some of our assumptions.
We have assumed the validity of a next to probe-brane approximation.
For this assumption to be valid, the mass of the portion of the brane
near the black hole horizon must be much smaller than the black hole
mass $\sigma r_h^3 \ll M$~\cite{Flachi06b}. 
This conditions translates to

\begin{equation}
\hat{\sigma} \ll \left( \frac{M}{M_\mathrm{P}} \right)^\frac{D-6}{D-3}\, .
\end{equation}

\noindent
So the next to probe-brane approximation will be valid until $M \sim
M_\mathrm{P}$, ie.\ if $\hat{\sigma} \ll 1$.

Energy-momentum conservation in four dimensions is a result of
translational invariance in four-dimensional spacetime.
The three-brane breaks the translational invariance in the extra
dimensions and hence momentum in these directions need not be conserved
in interactions between the bulk and brane states.
Nevertheless, we have assumed energy-momentum conservation between the
gravitons and the black hole in all the dimensions.

We have generated black holes with mass well above the Planck scale in
order to work in the regime of classical gravity.
However, the black hole eventually decays down to the Planck scale
and quantum gravity probably becomes important.
The effects of black hole recoil are most significant near the Planck
scale. 
Thus black holes may behave differently than we have depicted, but we
might expect the concepts of black hole recoil and missing energy to
remain unchanged.

The studies presented here do not include parton fragmentation,
hadronization, decay, detector effects, or backgrounds.   
Including these effects is likely to change the missing energy and mass
distributions for black hole events.
However, it is anticipated that the qualitative results for the missing
energy, black hole mass, and cross section will remain unchanged.

We draw two conclusions from our study of missing energy in black hole
evaporation: 
1) black holes will leave the brane less than 1/3 of the time
at the LHC, and
2) for very weak brane tensions, the irreducible acceptance for the
detection of black holes above 5~TeV can be as low as 87\%.

We have only studied the missing energy in black hole evaporation.
Missing energy from graviton emission during and shortly after black
hole formation could be more significant.
This graviton emission will have to be better understood before the
cross section can be measured and the Planck scale and number of
dimensions determined.   

\section*{Acknowledgments}

This work was supported in part by the Natural Sciences and Engineering
Research Council of Canada, and the Faculty of Science University
College London. 
I thank the Flower Kings for inspiration.

\bibliographystyle{JHEP}
\bibliography{recoil}

\end{document}